\documentclass[aps,pra]{revtex4}
\usepackage{graphicx}

\begin{document}

\title{Secure direct bidirectional communication protocol using
the Einstein-Podolsky-Rosen pair block \\
\thanks{*Email: zhangzj@wipm.ac.cn }}

\author{Z. J. Zhang and  Z. X. Man \\
{\normalsize Wuhan Institute of Physics and Mathematics, Chinese
Academy of Sciences, Wuhan 430071, China } \\
{\normalsize *Email: zhangzj@wipm.ac.cn }}

\date{\today}
\maketitle

\begin{minipage}{380pt}
In light of Deng-Long-Liu's two-step secret direct communication
protocol using the Einstein-Podolsky-Rosen pair block [Phys. Rev.
A {\bf 68}, 042317 (2003)], by introducing additional local
operations for encoding, we propose a brand-new secure direct
communication protocol, in which two legitimate users can
simultaneously transmit their different secret messages
to each other in a set of quantum communication device.\\

PACS Number(s): 03.67.Hk, 03.65.Ud \\
\end{minipage}\\

Quantum key distribution (QKD) is an ingenious application of
quantum mechanics, in which two remote legitimate users (Alice and
Bob) establish a shared secret key through the transmission of
quantum signals. Much attention has been focused on QKD after the
pioneering work of Bennett and Brassard published in 1984 [1].
Till now there have been many theoretical QKDs [2-19]. They can be
classified into two types, the nondeterministic one [2-14] and the
deterministic one [15-19]. The nondeterministic QKD can be used to
establish a shared secret key between Alice and Bob, consisting of
a sequence of random bits. This secret key can be used to encrypt
a message which is sent through a classical channel. In contrast,
in the deterministic QKD, the legitimate users can get results
deterministically provided that the quantum channel is not
disturbed. It is more attractive to establish a deterministic
secure direct communication protocol by taking advantage of the
deterministic QKDs. However, different from the deterministic
QKDs, the deterministic secure direct communication protocol is
more demanding on the security. Hence, only recently a few of
deterministic secure direct protocols have been proposed
[15-16,19]. One of these protocols is Deng-Long-Liu's two-step
quantum direct communication protocol using the EPR pair block
[19]. It is provably secure and has a high capacity. However, this
deterministic secure direct protocol is also a
message-unilaterally-transmitted protocol as well as the protocols
in [15-16], i.e., two parties can not simultaneously transmit
their different secret messages to each other in a set of quantum
communication device. In general, convenient bidirectional
simultaneous mutual communications are very useful and usually
desired. In this paper in light of Deng-Long-Liu's communication
protocol, by introducing additional local operations for encoding,
we propose a secure direct bidirectional communication protocol,
in which two legitimate users can simultaneously transmit their
secret messages to each other in a set of quantum communication
device.

Let us start with a brief description of the two-step protocol.
Alice prepares an ordered $N$ EPR photon pairs in state $|\Psi
\rangle _{CM}=|\Psi ^{-}\rangle =(|0\rangle _{C}|1\rangle
_{M}-|1\rangle _{C}|0\rangle _{M})/\sqrt{2}$ for each and divides
them into two partner-photon sequences $[C_1, C_2, \dots, C_N]$
and $[M_1, M_2, \dots, M_N]$, where $C_i$ $(M_i)$ stands for the
$C$ ($M$) photon in the $i$th photon pair. Then she sends the $C$
photon sequence to Bob. Bob chooses randomly a fraction of photons
in the $C$ sequence and tells Alice publicly which photons he has
chosen. Then Bob chooses randomly one of two measurement bases
(MB), say $\sigma_z$ or $\sigma_x$, to measure the chosen photons
and tells Alice which MB he has chosen for each and the
corresponding measurement result. Alice uses the same MB as Bob to
measure the corresponding partner photons in the $M$ sequence and
checks with Bob's result. Their results should be anticorrelated
with each other provided that no eavesdropping exists [16]. If Eve
is in the line, they have to discard their transmission and abort
the communication. Otherwise, Alice performs the unitary
operations on the unmeasured photons in the $M$ sequence to encode
her messages according to the following correspondences: $U_0
\leftrightarrow 00$; $U_1 \leftrightarrow 01$; $U_2
\leftrightarrow 10$; $U_3 \leftrightarrow 11$, where
$U_0=I=|0\rangle \langle 0|+|1\rangle \langle 1|$, $U_1=\sigma
_{z}=|0\rangle \langle 0|-|1\rangle \langle 1|$, $U_2=\sigma
_{x}=|1\rangle \langle 0|+|0\rangle \langle 1|$, $U_3=i\sigma
_{y}=|0\rangle \langle 1|-|1\rangle \langle 0|$. Then Alice sends
Bob the photons on which unitary operations has been performed .
After Bob receives the photons, he perform Bell-basis measurement
on each with its partner photon in the initial pair. Since
$U_0|\Psi ^{-}\rangle=|\Psi ^{-}\rangle$, $U_{1}|\Psi
^{-}\rangle=|\Psi ^{+}\rangle$, $U_{2}|\Psi ^{-}\rangle=|\Phi
^{-}\rangle$ and $U_{3}|\Psi ^{-}\rangle)=|\Phi ^{+}\rangle$, Bob
can extract Alice's encoding according to his measurement results.
By the way, in Alice's second transmissions, a small trick like
message authentification is used by Alice to detect on Eve's
attack without eavesdropping. In [19], the security of the
two-step protocol is proven.

Let us turn to our protocol. We only revise the two-step protocol
in a subtle way, however, the function of the protocol is changed
excitedly, i.e., two legitimate users can transmit simultaneously
their different secret messages to each other in a set of quantum
communication device. When Bob receives the photons on which Alice
has performed unitary operations to encode her messages, he does
not perform the Bell-basis measurements on each with its partner
photon in the initial pair at once but carry out a unitary
operation (i.e., $ U_{0},U_{1},U_{2}$ or $U_{3})$ on anyone photon
of the initial pair to encode his own message. After his unitary
operations Bob performs the Bell-basis measurements on the photon
pairs and publicly announces his measurement results. Since Bob
knows which unitary operation he has performed on one photon of
each pair, he can still extract Alice's encodings according to his
measurement results (See Table 1). Meanwhile since Alice knows
which unitary operation she has performed on one photon of each
pair, also she can extract Bob's encodings according to Bob's
public announcements of his measurement results (See Table 1). So
far we have proposed a deterministic direct bidirectional communication protocol. \\

\begin{minipage}{370pt}
\begin{center}
Table 1.  Corresponding relations among Alice's, Bob's unitary
operations (i.e., the encoding bits) and Bob's Bell measurement
results on the photon pair.  Alice's (Bob's)
unitary operations are listed in the first column (line).\\
\begin{tabular}{ccccc} \hline
& $U_{0}(00)$ & $U_{1}(01)$ & $U_{2}(10)$ & $U_{3}(11)$ \\ \hline
$U_{0}(00)$ & $|\Psi ^{-}\rangle $ & $|\Psi ^{+}\rangle $ & $|\Phi
^{-}\rangle $ & $|\Phi ^{+}\rangle $ \\
$U_{1}(01)$ & $|\Psi ^{+}\rangle $ & $|\Psi ^{-}\rangle $ & $|\Phi
^{+}\rangle $ & $|\Phi ^{-}\rangle $ \\
$U_{2}(10)$ & $|\Phi ^{-}\rangle $ & $|\Phi ^{+}\rangle $ & $|\Psi
^{-}\rangle $ & $|\Psi ^{+}\rangle $ \\
$U_{3}(11)$ & $|\Phi ^{+}\rangle $ & $|\Phi ^{-}\rangle $ & $|\Psi
^{+}\rangle $ & $|\Psi ^{-}\rangle $\\ \hline
\end{tabular}\\
\end{center}
\end{minipage}\\
\vskip 1cm

Let us discuss the security of our protocol. Before Bob's
announcement, the present protocol is only nearly same with the
two step protocol due to the additional unitary operations of Bob.
However, since all the photons are in Bob's hand, Eve can not know
which unitary operation Bob has performed at all. In fact, in this
case the essence of our protocol is the two-step protocol. Hence
it is secure for Bob to get the secret message from Alice
according to his Bell-basis measurements. Although later Bob
publicly announces his Bell measurement results, because he has
performed unitary operations which Eve can not know at all, Eve
still can not know which unitary operations Alice has ever
performed. Hence, it is still secure for Bob to get the secret
message from Alice via our protocol. Now that Eve can not know
which unitary operations Alice has performed and Bob publicly
announces his measurement results, Alice can securely know which
unitary operation Bob has ever performed, i.e., she can extract
securely Bob's encodings. Hence the present quantum dense coding
protocol is secure against eavesdropping. As for Eve's attack
without eavesdropping, we can also adopt the strategy as the trick
in [19] to detect it.

To summarize, we have proposed a deterministic secure direct
bidirectional communication protocol by using the
Einstein-Podolsky-Rosen pair block. In this protocol two
legitimate users can simultaneously transmit their different
secret messages to each other in a set of quantum communication
device.

This work is supported by the National Natural Science Foundation
of China under Grant No. 10304022. \\

\noindent [1] C. H. Bennett and G. Brassard, in {\it Proceedings
of the IEEE International Conference on Computers, Systems and
Signal Processings, Bangalore, India} (IEEE, New York, 1984),
p175.

\noindent[2] A. K. Ekert, Phys. Rev. Lett. {\bf67}, 661 (1991).

\noindent[3] C. H. Bennett, Phys. Rev. Lett. {\bf68}, 3121
 (1992).

\noindent[4] C. H. Bennett, G. Brassard, and N.D. Mermin, Phys.
Rev. Lett. {\bf68}, 557(1992).

\noindent[5] L. Goldenberg and L. Vaidman, Phys. Rev. Lett.
{\bf75}, 1239  (1995).

\noindent[6] B. Huttner, N. Imoto, N. Gisin, and T. Mor, Phys.
Rev. A {\bf51}, 1863 (1995).

\noindent [7] M. Koashi and N. Imoto, Phys. Rev. Lett. {\bf79},
2383 (1997).

\noindent[8] W. Y. Hwang, I. G. Koh, and Y. D. Han, Phys. Lett. A
{\bf244}, 489 (1998).

\noindent[9] P. Xue, C. F. Li, and G. C. Guo,  Phys. Rev. A
{\bf65}, 022317 (2002).

\noindent[10] S. J. D. Phoenix, S. M. Barnett, P. D. Townsend, and
K. J. Blow, J. Mod. Opt. {\bf42}, 1155 (1995).

\noindent[11] H. Bechmann-Pasquinucci and N. Gisin, Phys. Rev. A
{\bf59}, 4238 (1999).

\noindent[12] A. Cabello, Phys. Rev. A {\bf61},052312 (2000);
{\bf64}, 024301 (2001).

\noindent[13] A. Cabello, Phys. Rev. Lett. {\bf85}, 5635 (2000).

\noindent[14] G. P. Guo, C. F. Li, B. S. Shi, J. Li, and G. C.
Guo, Phys. Rev. A {\bf64}, 042301 (2001).

\noindent[15] A. Beige, B. G. Englert, C. Kurtsiefer, and
H.Weinfurter, Acta Phys. Pol. A {\bf101}, 357 (2002).

\noindent[16] Kim Bostrom and Timo Felbinger, Phys. Rev. Lett.
{\bf89}, 187902 (2002).

\noindent[17] G. L. Long and X. S. Liu, Phys. Rev. A {\bf65},
032302 (2002).

\noindent[18] F. G. Deng and G. L. Long, Phys. Rev. A {\bf68},
042315 (2003).

\noindent[19] F. G. Deng, G. L. Long, and X. S. Liu, Phys. Rev. A
{\bf68}, 042317 (2003).

\end{document}